\documentclass[10pt,journal,compsoc]{IEEEtran}
\usepackage{cite}
\usepackage{graphicx}
\usepackage{subcaption}
\usepackage{listings}
\usepackage{opera}
\usepackage{hyperref}
\usepackage{tikz}
\usetikzlibrary{mindmap,backgrounds,calc,shadows,decorations.fractals,shapes,arrows,decorations.pathmorphing,automata,positioning,fit}
\usepackage{pgfplots}
\usepackage{xcolor}
\definecolor{native}{HTML}{C8C8C8}
\definecolor{ifc}{HTML}{FDAE61}
\definecolor{allocation}{HTML}{680000}
\definecolor{lsm}{HTML}{3366FF}
\usepackage{resizegather}
\usepackage{url}

\let\llncssubparagraph\subparagraph
\let\subparagraph\paragraph
\usepackage[compact]{titlesec}
\let\subparagraph\llncssubparagraph

\usepackage{amsthm}
\newtheorem{definition}{Definition}

\begin{document}
\title{CamFlow: Managed data-sharing\\ for cloud services}

\author{Thomas F. J.-M. Pasquier,~\IEEEmembership{Member,~IEEE,}
		Jatinder Singh,~\IEEEmembership{Member,~IEEE,} 
       David Eyers,~\IEEEmembership{Member,~IEEE}
       and~Jean Bacon~\IEEEmembership{Fellow,~IEEE,}
\IEEEcompsocitemizethanks{\IEEEcompsocthanksitem Thomas F. J.-M. Pasquier, Jatinder Singh and Jean Bacon are with the Computer Laboratory, University of Cambridge, UK.\protect\\
E-mail: firstname.lastname@cl.am.ac.uk
\IEEEcompsocthanksitem David Eyers is with the Department of Computer Science, University of Otago, New Zealand.\protect\\
E-mail: dme@cs.otago.ac.nz
}\thanks{Manuscript received 31 Mar. 2015; revised 25 Aug. 2015; accepted 11 Sept. 2015; current version 6 Oct. 2015.
}
}

\IEEEcompsoctitleabstractindextext{
\begin{abstract}
A model of cloud services is emerging whereby a few trusted providers manage the underlying hardware and communications whereas many companies build on this infrastructure to offer higher level, cloud-hosted PaaS services and/or SaaS applications. 
From the start, strong isolation between cloud tenants was seen to be of paramount importance, provided first by virtual machines (VM) and later by containers, which share the operating system (OS) kernel.
Increasingly it is the case that {\em applications} also require facilities to effect {\em isolation and protection} of data managed by those applications. They also require
{\em flexible data sharing} with other applications, often across the traditional cloud-isolation boundaries; for example, when government, consisting of different departments, provides services to its citizens through a common platform. 

These concerns relate to the management of data.
Traditional access control is application and principal/role specific, applied at policy enforcement points, after which there is no subsequent control over where data flows; a crucial issue once data has left its owner's control by cloud-hosted applications and within cloud-services. 
Information Flow Control (IFC), in addition, offers system-wide, end-to-end, flow control based on the properties of the data. We discuss the potential of cloud-deployed IFC for \emph{enforcing} owners' data flow policy with regard to protection and sharing, as well as safeguarding against malicious or buggy software.
In addition, the audit log associated with IFC provides transparency and offers system-wide visibility over data flows. This helps those responsible to meet their data management obligations,  providing evidence of  compliance, and aids in the identification of policy errors and misconfigurations. 
We present our IFC model and describe and evaluate our IFC architecture and implementation (CamFlow). This comprises an OS level implementation of IFC with support for application management, together with an IFC-enabled middleware. 

 \end{abstract}

\begin{keywords}
Compliance, Security, Audit, Cloud Computing, Information Flow Control, Middleware, PaaS
\end{keywords}}

\maketitle

\IEEEdisplaynotcompsoctitleabstractindextext

\IEEEpeerreviewmaketitle

\section{Introduction and Motivation}
\label{sec:introduction}

\IEEEPARstart{A}{ model} of cloud services is emerging whereby a few trusted providers manage the underlying hardware and communications infrastructure---datacenters with worldwide replication to achieve high data integrity and availability at low latency. 
Many companies build on this infrastructure to offer higher level cloud services, for example Heroku is a PaaS built on Amazon's EC2, above which SaaS offerings can be built (\eg the LIFX smart lightbulb cloud service on top of the Heroku platform).
From the start, protection was a paramount concern for the cloud as infrastructure is shared between tenants. Strong tenant isolation was provided by means of totally separated virtual machines (VMs) \cite{barham2003a, kivity2007kvm}
and more recently, isolated containers have been provided that share a common OS kernel \cite{bernstein2014containers}.

Increasingly, cloud-hosted applications may need not only protection (and isolation) from other applications but also have requirements for \emph{flexible data sharing}, often across VM and container boundaries. 
An example is the UK GCloud\footnote{\url{https://www.gov.uk/digital-marketplace}} initiative, a government platform designed to encourage small companies to provide cloud-hosted applications. These applications need to be composed and made to interoperate to support citizens' needs for online services. 
Similarly, the Massachusetts Open Cloud \cite{DesnoyersIC2E2015} is a marketplace (Open Cloud Exchange (OCX)) to encourage small business development.
Solutions are open and one may build on the services of another. The aim is to create a catalyst for the economic development of business clusters. 

End-users of cloud services still need to be assured that their data is protected from leakage to other parties by their cloud hosts, due to software bugs or misconfigurations, also safeguarded to the extent possible against insider attacks and external threats. 
But increasingly, they also need to be able to access their own data across applications and to share their data with others, according to the policies they specify. 
Containment mechanisms, such as VMs and containers, provide strong isolation between applications, but do not support these sharing requirements. 
The incorporation of cloud services within `Internet of Things' (IoT) architectures~\cite{mineraud2014} is another driver of the requirement for both protection and cross-application data sharing, given these IoT architectures' strong emphasis on (safe) interaction.
For example, a patient being monitored at home may store sensor-gathered medical data in the cloud and share it with selected carers,  medical practitioners, and medical research (big-data) repositories, via cloud-hosted and mediated services. Once data has left end-users' homes for cloud services, they need to be assured that it is only accessed as they specify. 

Traditional access control tends to be principal/role specific, and apply only within the context of a particular application\slash service. Controls are applied at policy enforcement points, after which there is no subsequent control over where data flows. Once data has left the direct control of its owner, for example, after being shared with others, 
it is difficult using traditional access controls to ensure and demonstrate that it is not leaked. 
If a leak is suspected, it often cannot be established whether this is a breach of confidentiality by a person or due to buggy or misconfigured cloud service software. 

Encryption offers protection by restricting access to {\em intelligible} data, even beyond the boundary of one's technical control. However, encryption hinders flexible, nuanced data sharing, in that key management (distribution, revocation) is difficult. 
Further, traceability is limited, as being mathematically based there
is generally no feedback as to when\slash where decryption occurs; and a compromised key or broken encryption scheme at any time in the future places data at risk. As such, it is important that data flows are managed and audited, even if data items are encrypted.

Although contracts exist between cloud providers and tenants, and cloud services are increasingly subject to regulation~\cite{millard2013cloud}, there is at present no way to establish that providers remain in compliance with these agreements and requirements. Also, there are often requirements that data should pass through certain processes, e.g., encryption or anonymisation. There is currently no clear mechanism to express such requirements and demonstrate they have been consistently enforced.

An approach to maintaining the association of data with policy is to use ``sticky policies'' \cite{stickyPol:2011}.
Here, owner-specified management constraints are attached to encrypted data.
Decryption is only allowed by parties accepting the management constraints and able to enforce them.
This forms the basis for establishing contractual relationships between data owners and service providers or other applications.
However, this approach requires trust in a (relatively large amount of) software. 
Further, the enforcement is either at too coarse a granularity or prohibitively expensive.
This is further explored in \S\ref{sec:background:sticky}.

As an alternative, Information Flow Control (IFC) augments traditional access control by offering continuous, system-wide, end-to-end flow control based on properties of the data---for example, ``medical data may only be used for research purposes after going through consent checking and anonymisation''.
IFC allows \emph{security contexts} to be defined system-wide and guarantees non-interference between them. This is achieved by tags applied to entities (e.g., processes, files, database entries), inseparable from the entities they are associated with.
Every exchange of data between entities is verified against security-context-domain relationships created by the tags, thus allowing tight control over any subsequent transfers of the data.

In this paper we present CamFlow (\textbf{Cam}bridge \textbf{Flow} Control Architecture). 
We outline CamFlow's IFC model and implementation which comprises a new operating system (OS) level implementation of IFC as a Linux Security Module (LSM), with support for application management, together with an IFC-enabled middleware.
IFC tags are checked on OS system calls and on message passing by the middleware, to determine whether data flows are permissible. Log records can be made efficiently of all attempted flows, whether permitted or rejected, and this log provides a possible basis for audit, data provenance and compliance checking. By this means it can be checked whether application level policy has been enforced and whether cloud service provision has complied with contractual obligations. 

We argue that incorporating IFC into the underlying PaaS-provided OSs, as a small, trusted computing base
would greatly enhance the trustworthiness of cloud services, whether public or private, and hence all their hosted services/applications. 
Our evaluation shows that IFC would incur acceptable overhead and our IFC model is designed to ensure that application developers need not be aware of IFC, although some application providers may wish to take explicit advantage of IFC. We demonstrate the feasibility of our approach via an IFC-enabled framework for web services, see \S\ref{sec:example}.

\noindgras{Contributions:} 
Our main contribution is to demonstrate the feasibility of providing IFC as part of cloud software infrastructure and showing how IFC can be made to work end-to-end, system-wide. 
In addition to discussing the 'big picture', in this paper we also present a new kernel implementation of IFC and a new audit function. 
Our approach enables:
(1) protection of applications from each other (non-interference); 
(2) flexible, managed data sharing across isolation boundaries;
(3) prevention of data leakage due to bugs/misconfigurations;
(4) extension of access control beyond application boundaries;
(5) increased transparency, through detailed logs of information flow decisions.

\S\ref{sec:related} gives background in protection and IFC, then 
\S\ref{sec:model} presents the essentials of the CamFlow IFC model, with examples. 
\S\ref{sec:os} and \S\ref{sec:mw} 
describe our new OS-level implementation of IFC as a LSM and its integration via trusted processes with an IFC-enabled middleware, storage services, etc. 
\S\ref{sec:audit} emphasises that audit in IFC systems produces logs capable of being processed by `big-data' analytics tools. Audit is central to establishing provenance and for providers to demonstrate compliance with contract and regulation. 
\S\ref{sec:example} shows how standard web services are supported transparently by the CamFlow architecture: only a privileged application management framework need be aware of IFC and unprivileged application instances can run unchanged. In all cases, evaluation is included within the section. 
\S\ref{sec:conclusion} summarises, concludes and suggests future work.

\section{Background}
\label{sec:related}

We first define the scope of current isolation mechanisms, highlighting the need for flexible data sharing {\em at application-level granularity}, \ie where applications manage their own security concerns, as well as strong isolation between tenants and/or applications.
As an introduction to IFC we outline the evolution of IFC models. Related work on IFC implementation at the OS level and within distributed systems is given with the relevant sections. 
We end with a brief comparison of IFC with taint tracking (TT) and sticky policies.

\subsection{IFC Models}
\label{sec:related:models}

In 1976, Denning \cite{Denning:1976:LMS:360051.360056} proposed a Mandatory Access Control (MAC) model to track and enforce rules on information flow in computer systems.  
In this model, entities are associated with security classes. The flow of information from an entity $a$ to an entity $b$ is allowed only if the security class of $b$ (denoted $\underline{b}$) is equal to or higher than $\underline{a}$. 
This allows the \emph{no-read up, no-write down} principle of Bell and LaPadula \cite{bell73} to be implemented to enforce secrecy. By this means a traditional military classification \emph{public, secret, top secret} can be implemented.
A second security class can be associated with each entity to track and enforce integrity (quality of data); 
\emph{no read down, no write up}, as proposed by Biba~\cite{citeulike:3017234}. A current example might allow input of information from a government website in the $.gov.uk$ domain but forbid that from ``Joe's Blog''. 
Using this model we are able to control and monitor information flow to ensure data secrecy and integrity.

In 1997 Myers \cite{Myers97decentralized} introduced a Decentralised IFC model (DIFC)
that has inspired most later work. This model was designed to meet the changing needs of systems from global, static, hierarchical security levels to a more flexible system, able to capture the needs of different applications. In this model each entity is associated with two labels: a \emph{secrecy} label and an \emph{integrity} label, to capture respectively the privacy/confidentiality of the data and the reliability of a source of data. Each label comprises a set of tags, each of which represents some security concern. Data is allowed to flow if the security label of the sender is a subset of the label of the receiver, and conversely for integrity.

Implementations of a decentralised model akin to Myers' include a sensitive embedded system for BMW cars \cite{bouard2013practical} and XBook \cite{singh2009xbook} in a social media context. Our own model is described in \S\ref{sec:model}.
When implemented from the OS kernel level, applications running under IFC enforcement do not need to be trusted for the data management policy 
to be properly enforced \cite{Krohn:2007:IFC:1294261.1294293}.

\subsection{Protection via VMs and Containers}

Isolation of tenants in cloud platforms is through hypervisor-supported virtual machines \cite{barham2003a, kivity2007kvm} or OS-provided containers \cite{bernstein2014containers}. 
However, flexible sharing mechanisms are also required to manage data exchange
between applications contributing to more complex systems, or to achieve end-user goals. For example, government applications might access citizens' records for various purposes; a user's data from different applications might together contribute to evidence related to health or wellbeing. 

At present, the sharing of information between applications 
tends to involve a binary decision (i.e. to share or not), as for example in Google \emph{pods} (containers).\footnote{https://cloud.google.com/container-engine/docs/pods/}
Whole resources can be shared, but no control over data usage between applications is provided.
Furthermore, there are no means for preventing leakage outside of the mechanisms implemented by the individual applications\slash services.

Solutions have been proposed to provide intra-application 
sandboxes
(down to individual end-users) \cite{lee2013pibox}, but such schemes are difficult to scale, require changes in application logic, and still do not provide control beyond isolation boundaries (\ie again, loss of control once the data is shared).

IFC has been proposed to guarantee the proper usage of data by social network applications \cite{singh2009xbook}.
The aim is to provide purpose-based disclosure via IFC \cite{kumar2014realizing} between isolated components, thus guaranteeing that shared data can only be used for a well-defined and agreed-upon purpose. 

IFC is by no means proposed as a replacement for access control, VMs or containers, but rather as a complement to those techniques to provide flexible, managed data-sharing. IFC would allow tenants and end-users to maintain control (within an IFC-enforcing world) and define policy applying to their data consistently and beyond isolation and application borders.

\subsection{Taint Tracking (TT) Systems} 
Runtime, dynamic TT is similar to IFC but with less functionality. TT systems use one tag type ``taint'' instead of secrecy and integrity tags. Tags propagate with data and data flows may be logged.
An entity that inputs tagged data acquires the data's tag(s). Data flow constraints are only enforced at specified sink points, for example, when data 
attempts to leave a mobile phone \cite{Enck:2010:TIT:1924943.1924971}. Policy is applied at sink points such as preventing private, unanonymised or unencrypted data from flowing, or strictly controlling to where data may flow.

An example of TT used for integrity purposes is to taint data from untrusted sources, e.g., user input from a TCP stream in a web application environment, and enforce that it is sanitised before being processed \cite{papagiannis2011php}. 
This simple mechanism prevents injection attacks that plague badly designed web applications.
An example of TT used for confidentiality purposes is to taint sensitive information, e.g., a list of contacts in a mobile phone, and track it through this closed system \cite{Enck:2010:TIT:1924943.1924971}. Data leaving the system (i.e. the phone) is analysed to ensure it does not contain sensitive information. Data containing sensitive information should only leave to a number of closely controlled destinations, such as the cloud backup contact list. This approach aids the detection of malicious applications attempting to steal user-sensitive information and send it to third parties.
Equally, this type of concern can be captured through the use of IFC policies.

One concern with TT systems is that there is a gap in time between the occurrence of the issue (\eg a leak, an attack) and when it is detected \cite{Schwartz2010}
\ie problems become evident only when the tainted data reaches a sink (enforcement point). 
Depending on the degree of isolation between the different parts of the system, and the number of system components involved, this tainted data may have `contaminated' much of the system.
While this can be managed in smaller, closed environments, it is less appropriate for cloud services in general.
IFC policies present the clear advantage to \emph{prevent} problems as they occur and to stop their effects propagating to a potentially large part of the system.

Some argue that TT is simpler to use than IFC, and incurs lower overhead,
 but when the enforcement is systemic and the granularity identical the overheads are similar (compare \cite{Enck:2010:TIT:1924943.1924971} and the evaluation in \S\ref{sec:os} and \S\ref{sec:mw}).
Indeed, the complexity of verifying IFC policy (see \S\ref{sec:model}) is  comparable to the cost of propagating taint. 
For both techniques, most of the overhead comes from the mechanism for intercepting data exchange.

\subsection{Sticky Policies}
\label{sec:background:sticky}

IFC can be seen as a mechanism for enforcing policy; the labels associated with entities represent application policies. 
IFC is a simple, low-level mechanism.
Sticky policy approaches also consider the enforcement of data-bound policy, but at a higher-level.

Casassa-Mont \etal \cite{hpsticky} first introduced \emph{sticky policies}, which involves encrypting data along with a list of policies to be enforced on that data.
To obtain the decryption key from a Trusted Authority (TA), a party must agree to enforce the policies associated with the data.
This agreement may be considered as part of forming a contractual link between the data owner and the service provider.
Work has continued in the area~\cite{bandhakavi2006super, chadwick2008enforcing, casassa2014towards}. 
Sticky policies, typically enforced at the application-level, are generally more complex and heavyweight than the simple secrecy and integrity constraints of IFC. As such, sticky policies tend only to be enforced at particular points, \eg at administrative boundaries.
IFC on the other hand, as we show in \S\ref{sec:os:evaluation}, can be enforced continuously at a reasonable cost.
In \S\ref{sec:model}, we discuss how complex policies might be built from IFC labels.

Further, the sticky policy approach builds upon the trust established between the data owner, the TAs and services that use the data.
A non-compliant service could be black-listed, but only if and when a breach of agreement is detected and the TA updated.
Our IFC approach builds only upon the trust between the data owner and the cloud provider.
Services and applications running on top of the cloud provider platform need not be trusted.
We believe this to be a great improvement to the overall trustworthiness of the system.

\section{CamFlow-Model: IFC for the Cloud}
\label{sec:model}

\begin{figure}[!t]
\centering
  \includegraphics[width=\columnwidth]{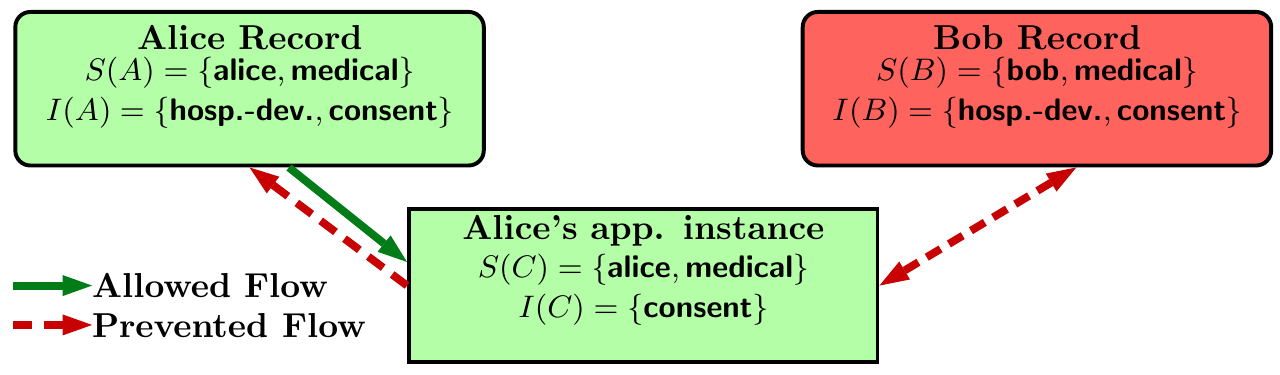}
  \caption{An allowed safe flow and prevented flows.} 
  \label{image:model:flow}
\end{figure}

\begin{figure*}[!t]
\centering
  \includegraphics[width=\textwidth]{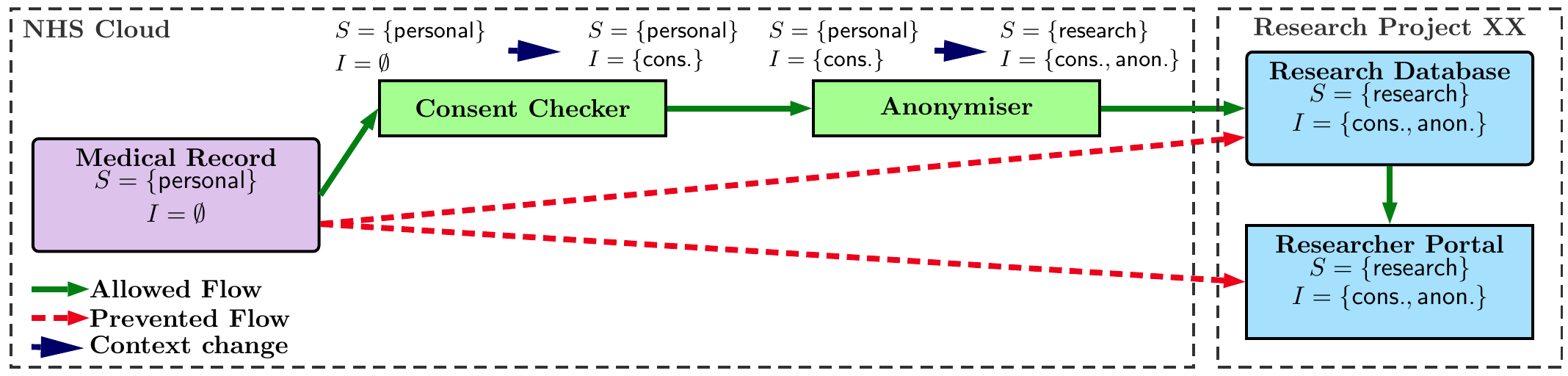}
  \caption{Medical data declassified and endorsed for research purposes.}
  \label{image:declass}
\end{figure*}

IFC operates to ensure that only permitted flows of information can occur, by enforcing data flow policy dynamically, end-to-end, within and across applications\slash services.
\emph{Entities} to which IFC constraints are
applied can include 
a MapReduce worker instance \cite{akoush14mrlazy}, a file, a process, a database entry \cite{Schultz:2013:IDI:2465351.2465357}, etc.
In CamFlow, IFC is applied continuously, typically on every system call for an IFC-enabled OS, and on communication mechanisms for enforcement across applications\slash runtime environments. IFC policy should therefore be as simple as possible, to allow verification, human understanding and to minimise runtime overhead.
Indeed, there is no need for IFC to encapsulate every possible policy; rather, it augments other control mechanisms, and can help enforce their policies.

\subsection{Tags and Labels}
\label{sec:ifc:tags}

We define tags that are tokens, each representing some security concern over secrecy or integrity. The tag $\textsf{\small bob-private}$ could for example represent Bob's personal data. We associate every entity in the system with two {\em labels} (sets of tags): an entity $A$ has a secrecy label $\secrecy{A}$ and an integrity label $\integrity{A}$. The state of these labels is the \emph{security context} of the entity.
The power of IFC is that it guarantees non-interference between security contexts \cite{von2004information, hedin2011perspective}.

\noindgras{Example -- secrecy: } 
Suppose a patient, Bob is discharged from hospital to be medically monitored  at home. The data streams from his sensors are transferred to a cloud service and are to be shared with his medical team at the hospital. The data items from his devices are tagged with $\textsf{\small medical, bob}$ in their secrecy labels.

\noindgras{Example -- integrity: }
The cloud-based home monitoring support service needs to be assured that the data it receives is 
from a hospital-issued device. 
Each sensing device is checked and issued with the tag $\textsf{\small hospital-device}$ in its integrity label.

Fig. \ref{image:model:flow} illustrates information flow constraints being applied over both secrecy and integrity dimensions.

\subsection{Decentralised Privileges and Security Contexts}
\label{sec:ifc:privileges}

In decentralised IFC (DIFC) any active entity can create \emph{new} tags. Tag creation is typically carried out by application managers when setting up application instances.
When an active entity creates a new tag either for secrecy or integrity, this process is given the corresponding privilege to add and remove the tag to its secrecy or integrity label respectively. If an active entity $A$ has a privilege to add $t$ to its secrecy label, we denote this $t \in \privilege{+}{S}{A}$, and to remove $t$ from its secrecy label: $t \in \privilege{-}{S}{A}$ (and similarly $\privilege{+}{I}{A}$ and $\privilege{-}{I}{A}$ are the privileges for integrity). 
An active entity may therefore have four privilege sets in addition to its security context.
Application managers will normally set up application instances in security contexts, without the privileges to change them. An example is given in \S\ref{sec:example}.

\subsection{Creating a New Entity}
\label{sec:ifc:create}

We define $A \Rightarrow B$ as the operation of the entity $A$ creating the entity $B$. An example is creating a process in a Unix-style OS by {\small \textsf{clone}}. We have the following rules for creation:
\begin{equation*}
\text{if }\create{A}{B}\text{, then}\begin{cases}
    S(B) := S(A) \\
    I(B) := I(A)
  \end{cases}
\label{sec:ifc_model:ifc_rule3}
\end{equation*}
That is, the created entity inherits the security context of its creator. These rules force the creating entity to explicitly change its security context to that required for the entity to be created. We motivate this below in \S\ref{sec:ifc:safe_label}.
Note that only labels pass to the created entity; privileges 
have to be passed explicitly.

\subsection{Security}
\label{sec:ifc:security}

The purpose of IFC models is to regulate flows between entities, and effect label changes and privilege delegation.

\begin{definition}
\label{def:safe_system}
A system is secure in the CamFlow IFC model if and only if all allowed messages are safe (Definition \ref{def:safe_message}), all allowed label changes are safe (Definition \ref{def:label_change})  and all privilege delegation is safe (Definitions \ref{def:privilege_deleg} and \ref{def:coi}).
\end{definition}

\subsubsection{Information Exchange}
\label{safemsg}

IFC prevents data leakage by controlling the exchange of information. We follow the classic pattern for IFC-guaranteed secrecy (\emph{no read up, no write down} \cite{bell73}) and integrity (\emph{no read down, no write up} \cite{citeulike:3017234}).

\begin{definition}
\label{def:safe_message}
A flow of information $\flow{A}{B}$ is safe if and only if:
\begin{equation*}
 \flow{A}{B} \text{, iff  }
   \{ S(A) \subseteq S(B) \wedge I(B) \subseteq I(A) \}
\end{equation*}
\end{definition}

\noindgras{Example -- secrecy enforcement: } 
Consider our example of patient monitoring after discharge from hospital, where the patient's devices are tagged with $\textsf{\small medical, bob}$ in their secrecy labels. In order for the cloud service to be able to receive this data it must also include the tags $\textsf{\small medical, bob}$ in its secrecy label.
Therefore an application instance accessing Bob's medical data must be labelled as such. In \S\ref{sec:example} we describe how applications can be designed to meet such requirements.
 
\noindgras{Example -- integrity enforcement: } 
The cloud-based home monitoring support service needs to be assured that the data it receives is from a hospital-issued device. To achieve this, the service has an integrity tag $\textsf{\small hospital-issued}$ in its integrity label and will only accept data from devices with tags $\textsf{\small hospital-issued}$.

\subsubsection{Label Change} 
\label{sec:ifc:safe_label}

Under the above constraints, information flows are restricted to equal or increasing secrecy constraints and equal or decreasing integrity constraints.
However, data may undergo transformations and/or checks that change its security properties.
For example, moving data through an anonymisation engine renders the data less sensitive, so less strict secrecy constraints can apply to the anonymised output. 
In the integrity dimension, data may go through a validation process on input, thus becoming more trustworthy.
In CamFlow 
only the process itself is able to change its secrecy and integrity labels, which requires the appropriate privileges and must be explicitly requested.

\begin{definition}
\label{def:label_change}
A label change noted $\contextchange{A}{A'}$ is safe if and only if for a label X (either S or I) and a tag $t$:
\begin{equation*}
\begin{multlined}
X(A'):=X(A)\cup\{t\} \text{ if } t \in P_X^+(A) \\
\text{OR}   \\
X(A'):=X(A) \setminus \{t\} \text{ if } t \in P_X^-(A) \\
\end{multlined}
\label{sec:ifc_model:ifc_rule4}
\end{equation*}
\end{definition}

Declassifiers and endorsers are the entities with the privileges to perform security context transformations. Declassifiers change the secrecy properties and endorsers change the integrity properties.

\noindgras{Example -- declassification: } 
A medical record system is held in a private cloud. Research datasets may be created from these records, but only from records where the patients have given consent. Also, only anonymised data may leave the private protected environment. We assume a health service approved anonymisation procedure. Fig. \ref{image:declass} shows the anonymiser inputting data tagged as $\textsf{\small personal}$ and declassifying the data by outputting data with secrecy tag 
$\textsf{\small research}$. 

\noindgras{Example -- endorsement: } 
In the same example, the Research Database is on a public cloud and may only receive research data  tagged with $\textsf{\small consent, anon}$ in its integrity label. In the private cloud we see a process that selects appropriate records for specific research purposes, checks for patient consent and adds the tag $\textsf{\small consent}$ to the integrity label of its output. The anonymiser process can only input data with this tag; it anonymises the data and outputs data with the tag  $\textsf{\small anon}$ in its integrity label.

Some previous work \cite{Krohn:2007:IFC:1294261.1294293, porter2014practical} allows implicit \emph{declassification} and \emph{endorsement}. That is, if an active entity has the privilege to declassify/endorse and the privilege to return to its original state (i.e. for declassification\slash endorsement over $t$ the entity has privilege $t^-$ and $t^+$), the declassification/endorsement may occur implicitly without the need for the entity to make the label changes. 
We believe that this could in practice lead to 
\emph{unintentional} data disclosure. Suppose an entity has the privilege to declassify top-secret information. The requirement for explicit label change makes it unlikely that the entity will send such data accidentally to an unintended recipient. 
Our model has stronger constraints that require endorsement and declassification operations to be programmed \emph{explicitly}.

\subsubsection{Privilege delegation} 
\label{sec:model:delegation}

An entity is only able to delegate a privilege it owns. 

\begin{definition}
\label{def:privilege_deleg}
A privilege delegation is safe if and only if $t \in P_X^\pm(A)$.
\end{definition}

\subsection{Conflict of Interest}
\label{sec:model:coi}

In CamFlow alone among IFC systems, privilege delegation is further restricted by Conflict of Interest (CoI) (or Separation of Duty (SoD)) enforcement. The receiving entity A, must not be put in a situation where it would break a CoI constraint. By this means, an application manager is prevented from creating an application instance with access to conflicting data.

\begin{definition}
\label{def:coi}
An entity $A$ does not violate a CoI $C$ if and only if:
\small
\begin{equation*}
\Big|\Big(S(A) \cup I(A) \cup P^+_S(A) \cup P^+_I(A) \cup P^-_S(A) \cup P^-_I(A)\Big) \cap C\Big| \leq 1
\label{sec:ifc_model:ifc_rule6}
\end{equation*}
\normalsize
\end{definition}

\noindgras{Example -- conflict of interest: }
A CoI might arise when data relating to competing companies is available in a system. In a hospital context, this might involve the results of analysis of the usage and effects of drugs from competing pharmaceutical companies. The companies might agree to this analysis only if their data is guaranteed to be isolated, i.e. not leaked to other companies. 

The hospital may be participating in drug trials and want to ensure that information does not leak between trials: 
suppose a conflict is 
$C=\{ \textsf{\small Pfizer}, \textsf{\small GSK}, \textsf{\small Roche}, ...\}$ 
and some data (e.g. files) are labelled $\textsf{\small PfizerData}[S=\{ \textsf{\small Pfizer}\}, I=\emptyset]$ and $\textsf{\small RocheData}[S=\{\textsf{\small Roche}\}, I=\emptyset]$. 
The CoI described ensures that it is not possible for a single entity (e.g. an application instance) to have access to both $\textsf{\small RocheData}$ and $\textsf{\small PfizerData}$ either simultaneously or sequentially, i.e. enforcing that Roche-owned data and Pfizer-owned data are processed in isolation.

\vspace{1.5mm}
The next sections describe the CamFlow platform that enforces the IFC constraints described.

\section{OS Enforcement}
\label{sec:os}
\begin{figure}[t]
\centering
  \includegraphics[width=\columnwidth]{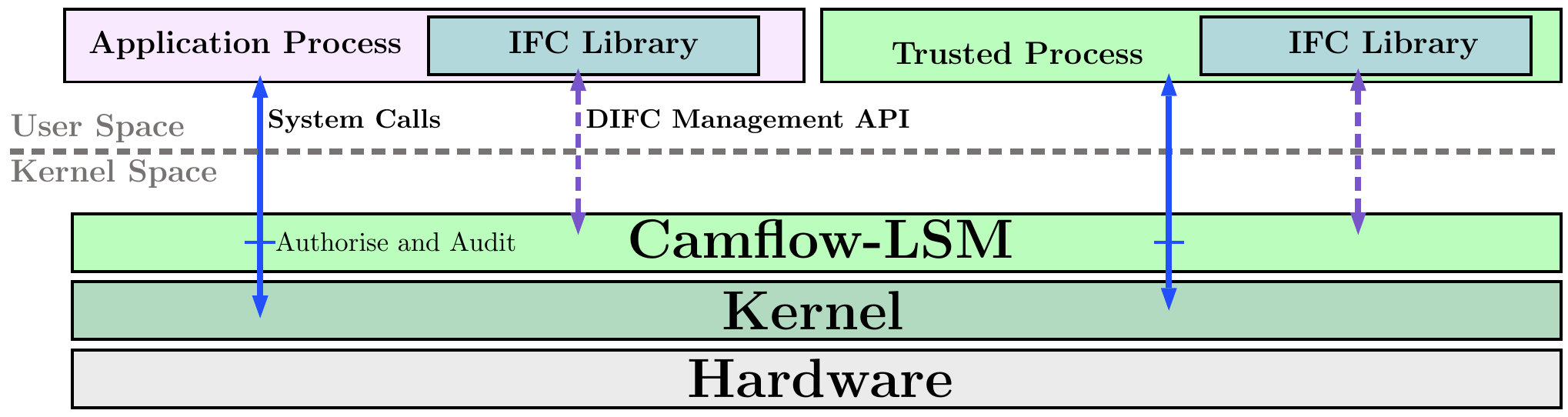}
\caption{
The interactions of the IFC Security Module (LSM) and a Trusted Process within an OS.
}
  \label{image:lsmarch}
\end{figure}

At the heart of the architecture  
is a minimal kernel module dedicated solely to OS-level IFC enforcement. 
The module is trusted to enforce IFC, transparently, across all flows between entities within the OS. User space processes can directly interact with the kernel module, \eg to delegate privileges (\S\ref{sec:ifc:security}) through a pseudo-file system, abstracted through a high level API. 
Higher level considerations and policies can be managed through specifically defined Trusted Processes (see \S\ref{sec:os:trusted}).
The local machine architecture is represented in \fig\ref{image:lsmarch}.

Note that IFC operates alongside and complements other security technologies. It is not a cloud security panacea; challenges regarding 
covert and side channels, and direct access to hardware by an attacker
remain, as they do for systems in general.
There are approaches that can help address these security threats, but many are highly disruptive (\eg synchronisation approaches to reducing timing channels) and are infrequently used.
Other threats may be easier to mitigate and solutions may be used when appropriate (\eg on-disk encryption).

\subsection{CamFlow-LSM} 
Our kernel module, {\em CamFlow-LSM}, is implemented as a  Linux Security Module (LSM) \cite{wright2003linux}.
Although our work is Linux-specific, a similar approach could be used on any system providing LSM-like security hooks.
Unlike other DIFC OS implementations \cite{Krohn:2007:IFC:1294261.1294293, porter2014practical} our kernel patch
is self-contained, strictly limited to the security module, does not modify any existing system calls and follows LSM implementation best practice.
This allows, among other things, LSM stacking \cite{quaritsch2004linux, schaufler2014} and coexistence with other security modules such as \eg SELinux \cite{smalley2001implementing} or AppArmor \cite{bauer2006paranoid} and complements their MAC enforcement with decentralised information flow policies.

We assume that the rest of the kernel can be trusted and does not interfere with the IFC enforcement mechanism.
LSM system hooks have been statically and dynamically verified \cite{edwards2002runtime, jaeger2004consistency, ganapathy2005automatic}, and our implementation inherits from LSM the formal assurance of IFC's correct placement on the path to any controlled kernel object.
This is sufficient to guarantee that we control flow and record audit on any operation on a controlled kernel object.

Since applications running on SELinux \cite{smalley2001implementing} or AppArmor \cite{bauer2006paranoid} need not be aware of the MAC policy being enforced, we see no reason to force applications running on an IFC system to be aware of IFC; only those performing declassification or endorsement operations are necessarily aware. 
This implementation choice is important; cloud providers can incorporate IFC without requiring changes in the software deployed by tenants. Alternatively, applications that wish to manage their own IFC constraints can declare policy through a pseudo-filesystem (as is typical for LSMs) abstracted by a user space library  and enforced transparently by the IFC mechanism.

The LSM framework calls security hooks when access to a kernel object is attempted.
Security metadata can be associated with kernel objects and is used by the LSM module to make access decisions.
Tags and privileges are represented by 64-bit opaque nonces associated with kernel objects such as processes, inodes, files, shared memory objects, messages \etc
On interaction between kernel objects, CamFlow-LSM security hooks are called to enforce data-flow policy (\S\ref{safemsg}) or propagate tags on entity creation (\S\ref{sec:ifc:create}) as appropriate.

Only active entities (processes) have mutable labels and privileges, all other (passive) entities have immutable labels and no privileges.

Privileges are allocated by the kernel and owned by the creating process (any process can create tags and the associated privileges in a decentralised fashion).
Privileges can be passed to other processes, users or groups, CamFlow-LSM verifying that constraints on privilege delegation (\S\ref{sec:model:delegation}) and conflict of interest (\S\ref{sec:model:coi}) are not violated.
A process
can add or remove a tag from its label if it owns the appropriate privilege (following IFC constraints described in \S\ref{sec:ifc:safe_label}), if the current user owns the privilege or if the current group owns the privilege.
How tags are shared and managed must be considered with care when designing an application and the system 
must be administered accordingly.

\subsubsection{Checkpointing and Restoration}
\label{sec:os:checkpoint}

Checkpointing a process involves halting its execution, allowing it to be restarted at a later stage, and enabling migration, \eg \cite{egwutuoha2013survey}.
LSM state is normally saved and restored by the checkpointing system, e.g. \cite{laadan2010linux}, and our module further exports an API to more efficiently serialise and restore security context.

Furthermore, self-checkpointing and restoring the previous state of a process, has been demonstrated \cite{niu2013efficient} to be a beneficial feature for IFC systems. 
This is particularly useful for processes serving requests.
In such a scenario the state of the process is saved after initialisation. 
When a request is received, the serving process sets itself up in the security context appropriate to serve the request.
After the request is served (or a series of requests if the system is session-based as described in \S\ref{sec:example}), the process restores its memory state and security context to what they were immediately after initialisation. This improves performance and prevents data leaks between security contexts.

\subsubsection{OS Evaluation}
\label{sec:os:evaluation}

We tested the CamFlow-LSM module on Linux Kernel version 3.17.8 (01/2015) from the Fedora distribution.\footnote{It is not feasible to provide a comparison with the Laminar implementation~\cite{porter2014practical}, that is closest in technical terms to our work, as the implementation available \url{https://github.com/ut-osa/laminar} is for an obsolete kernel version 2.6.22 (07/2007).}
The tests are run on an Intel 2.2Ghz i7 CPU and 6GiB RAM machine.

\begin{figure}[t]
\centering
\begin{tikzpicture}
\begin{axis}[
    xbar stacked,
    legend style={
    legend columns=4,
        at={(xticklabel cs:0.5)},
        anchor=north,
        draw=none
    },
    ytick=data,
    axis y line*=none,
    axis x line*=bottom,
    tick label style={font=\footnotesize},
    legend style={font=\footnotesize},
    label style={font=\footnotesize},
    xtick={0,10,20,30,40,50,60,70,80,90},
    width=\columnwidth,
    bar width=3mm,
    xlabel={Time in ms},
    yticklabels={sys\_pipe, sys\_write, sys\_read, sys\_clone},
    xlabel={$\mu$s},
    xmin=0,
    xmax=100,
    area legend,
    y=7mm,
    enlarge y limits={abs=0.625},
]
\addplot[allocation,fill=allocation] coordinates
{(4.821,0) (0,1) (0,2) (5.850, 3)};
\addplot[ifc,fill=ifc] coordinates
{(9.381,0) (3.441,1) (3.196,2) (27.355, 3)};
\addplot[lsm,fill=lsm] coordinates
{(1.229,0) (1.119,1) (1.028,2) (1.120, 3)};
\addplot[native,fill=native] coordinates
{(72.736,0) (61.981,1) (39.982,2) (57.48, 3)};
\legend{dyn. label, IFC, LSM, Native}
\end{axis}  
\end{tikzpicture}
  \caption{
  Overhead introduced into the OS by CamFlow LSM (x-axis time in $\mu$s).
    }
  \label{image:os:eval}
\end{figure}
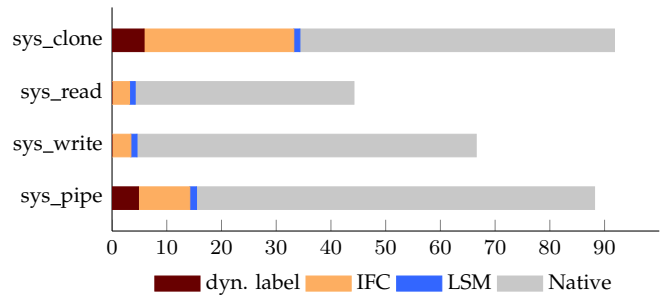

Measurements are done using the Linux tool \textsf{\small ftrace} \cite{bird2009measuring} to provide a microbenchmark. 
Two processes read from and write to a pipe respectively. Each has 20 tags in its security label, substantially more than we have seen a need for in current use cases. We measure the overhead induced by: 
creating a new process (\textsf{\small sys\_clone}), creating a new pipe (\textsf{\small sys\_pipe}), writing to the pipe (\textsf{\small sys\_write}) and reading from the pipe (\textsf{\small sys\_read}). The results are given in Fig. \ref{image:os:eval}.

We can distinguish two types of induced overhead: verifying an IFC constraint (\textsf{\small sys\_read}, \textsf{\small sys\_write}) and allocating labels (\textsf{\small sys\_clone}, \textsf{\small sys\_pipe}).
The \textsf{\small sys\_clone} overhead is roughly twice that of \textsf{\small sys\_pipe} as memory is allocated dynamically for the active entity's labels and privileges. Recall that passive entities have no privileges.  
Overhead measurements for other system calls/data structures 
are essentially identical as they rely on the same underlying enforcement mechanism, and are not included. 

The CamFlow-LSM overhead is a few percent, see Fig. \ref{image:os:eval}.
We provide a build option that further improves performance by declaring labels and privileges with a fixed maximum size (by default, label size can increase dynamically to meet application requirements). This reduces the overhead of the system calls that create new entities (the dynamic label component in Fig. \ref{image:os:eval}). 
This is an acceptable trade-off as in practical scenarios, labels rarely exceed more than five tags.
However, for most applications, the overhead is imperceptible and lost in system noise; it is hard to measure without using kernel tools, as the variation between two executions may be greater than the overhead.

\subsection{Trusted Processes}
\label{sec:os:trusted}

\begin{figure*}[t]
\centering
  \includegraphics[width=\textwidth]{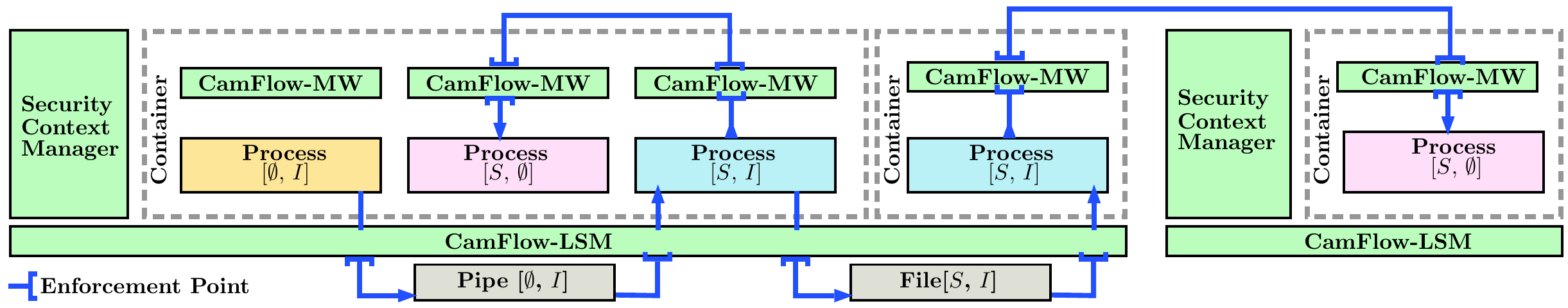}
  \caption{CamFlow Architecture: Labelled OS objects, trusted processes and communication middleware. 
    }
  \label{image:fullarchitecture}
\end{figure*}

The CamFlow-LSM is trusted to enforce IFC at the kernel level. Its functionality is minimal; strictly confined to the enforcement of IFC policies as described in \S\ref{sec:model}.
This guarantees easier maintainability and a system that is agnostic to higher level application requirements, thus minimising the constraints imposed on user-space application design.

We introduce the concept of a {\em trusted process}, that allows application/platform-specific concerns to be managed in user space by bypassing some LSM-enforced IFC constraints. For example, a trusted process might serve as a proxy for external connections, as in the Trusted IFC Gateway in the example in \S\ref{sec:example}, setting up and managing application components' labels. Trusted processes are used to interact with persistent storage (see \S\ref{sec:os:service}), for checkpointing and restoring processes (see \S\ref{sec:os:checkpoint}) and for managing inter-process and external communication (see \S\ref{sec:mw}).

Fig. \ref{image:fullarchitecture} shows OS instances running the CamFlow-LSM hosting a number of application processes, that may be grouped in containers. Each OS instance has a single trusted process (Security Context Manager) to manage its hosted processes' IFC labels and privileges. In addition, each process has an associated trusted middleware process to handle inter-process and inter-machine communication. Such communication may be within or between containers, OSs or clouds. 

In this example, $S$ represents a particular set of secrecy tags, and $I$  a particular set of integrity tags, both of which remain the same throughout. 
The application processes and other OS objects, such as pipes and files, are labelled $[S,I]$. The process labelled $[\emptyset,I]$ writes `public' data to a pipe, which is read by a process labelled $[S,I]$, assuming all the $I$ tags match correctly. 
Similarly, two processes are shown writing to and reading from a file.

The Security Context Manager maps between the kernel-level representation of tags (as 64-bit integers) and the representation of tags in user space. 
Within a cloud or other trusted environment, tags may be simple strings. When tags need to cross domain boundaries, e.g., when cloud services form part of a wider architecture, as in IoT, tags may need to be protected by cryptographic means (see \S\ref{sec:tags:represent}).

Trusted processes are either set up through static configuration, read at boot time by the CamFlow-LSM module, or created at runtime by another trusted process.
Trusted processes must either be managed by a trusted party (in our current approach the underlying infrastructure provider) and\slash or the code must be auditable and a means to verify the current version running on the platform must be provided (see \S\ref{sec:os:hardware}).

\subsection{Leveraging Hardware Roots of Trust}
\label{sec:os:hardware}

Incorporating IFC into cloud-provider OSs would enhance the trustworthiness of the platform. However, IFC only guarantees protection above the technical layer in which it is enforced. 
Recent hardware and software developments make it possible to attest that the software layers on which our platform runs have been audited.

The Trusted Platform Module (TPM) \cite{parno2008bootstrapping}, as used for remote attestation \cite{santos2009towards}, is one such hardware mechanism. 
TPM is used to generate a nearly unforgeable hash representing the state of the hardware and software of a given platform, that can be remotely verified.
Therefore, a company could audit the implementation of our IFC enforcement mechanism and ensure that our kernel security module, messaging middleware and the configuration they provide are indeed running on the platform. 
Any difference between the expected state of the software stack and the platform could be considered a breach of trust; such considerations can easily be embedded in the contractual obligations of the cloud provider.

TPM and remote attestation for cloud computing \cite{perez2006vtpm} are reaching maturity, with IBM rolling out an open source, scalable trusted platform based on virtual TPMs \cite{BergerIC2E2015}. 
Indeed, Berger \etal \cite{BergerIC2E2015} describe a mechanism allowing the TPM and remote attestation to be provided for virtual machine offerings and container-based solutions, covering the whole range of contemporary cloud offerings.
Furthermore, the approach not only allows the state of the software stack to be verified at boot time, but also during execution, and can thus prevent run-time modification of the system configuration.

\section{Cross-Machine Enforcement}
\label{sec:mw}

CamFlow-LSM operates to protect flows within the OS. However, it is also important that flows are protected across OS instances.

Generally, in order to guarantee flow constraints, only processes $P$ such that $S(P)=\emptyset$ and $I(P)=\emptyset$, \ie not subject to IFC constraints, are allowed to directly connect to or receive messages from  connections on remote OS (\eg through a socket). In order to connect to another machine, a process must either:
1) be able to declassify to change its security context to $S(P)=\emptyset$ and $I(P)=\emptyset$;
2) communicate through an intermediate trusted process.

As such, CamFlow contains an IFC-enabled, fully-featured messaging middleware (\textbf{CamFlow-MW}) to 
both facilitate communication and guarantee enforcement across machines.
For want of space, we only consider the middleware concepts relevant to IFC; details on the general middleware (as it was prior to IFC\slash CamFlow integration) can be found in \cite{debs14SBUStutorial}.
In short, the middleware supports strongly-typed messages; a range of interaction paradigms, including request-reply, broadcast, and streams; flexible resource discovery; and security mechanisms including access controls and encrypted communication. 
A particular feature is its support for dynamic reconfiguration based on event-driven policy. This simplifies both application development and deployment, as concerns can be abstracted and tailored to the particular environment, rather than embedded within application code.

The role of the middleware is to move towards continuous, end-to-end data flow management, such that IFC can be enforced {\em across} applications\slash machines (kernels).  
There 
is work on IFC enforcement across machines; however, these impose specific requirements, such as design-time considerations~\cite{sfaxi:CPE:CPE2807}, a particular language\slash runtime~\cite{yoshihama}, or  constraints on system architecture\slash implementation~\cite{cheng12:_abstr_usabl_infor_flow_contr_aeolus}. 
In contrast, we integrate IFC functionality into the general, fully featured distributed systems middleware mentioned above (see~\cite{singh2014ic2e}), to provide flexibility and be more generally applicable. 
We deliberately avoid imposing a structure on system design, instead integrating IFC functionality into the sort of communications infrastructure common to current enterprise and cloud systems.

\subsection{Remote Interactions}
\label{sec:tags:represent}

CamFlow-MW operates by associating a trusted process (see \S\ref{sec:os:trusted}) with an entity that seeks to communicate via the messaging system. Its fit within the broader architecture is depicted in \fig\ref{image:fullarchitecture}.
The process is responsible for handling the communication of messages, and enforces IFC based on the current runtime labels of the entity on whose behalf it operates.

It follows that  for IFC to be enforced across machines, tags require system-wide management, \ie throughout the cloud service. 
In \cite{singh2015riot} we proposed that the widely used and available X.509 certificates could be used.
The approach relies on public key certificates and attribute certificates~\cite{park2000binding}, 
to respectively identify the application associated with the CamFlow-MW instance and the tags associated with this application.

As part of establishing a connection, CamFlow-MW ensures that each entity authorises communication with the other, according to a local access control policy. Decisions are based on component metadata, the relevant authentication aspects secured through PKI (certificates).
Similarly, IFC policy must also be verified to ensure data flows are authorised according to IFC policy.
Attribute certificates provide cryptographic means to determine and verify the tags associated with the remote entity (see~\cite{singh2015riot}), on which policy can be enforced. If tags do not accord, the connection will not be established.

We also see potential for remote attestation, based on hardware integrity measures (see \S\ref{sec:os:hardware}), to be integrated into this authorisation phase, to ensure the remote machine operates a reliable IFC enforcement regime.

\subsection{Message-Level Enforcement}

CamFlow messages are strongly typed, where a {\em message type} is defined by a schema describing its set of attributes. 
For an instance of a message, an {\em attribute} consists of a name, type and value. 
 The support for IFC within messages is fine-grained, in that individual attributes within messages can also be labelled. These attribute labels introduce additional IFC constraints over and above those already applying to the entity, \ie as recognised by the kernel-LSM, and validated on connection establishment.

Labels can be defined within message type schema, which sets the attributes' IFC labels for all message instances of the type. These labels cannot be changed by entities dealing in such messages, and the entities must hold the requisite labels to interact with the attributes. 
Otherwise, the entity producing/publishing a message can set the security labels for the attributes (for those not predefined), if the entity holds the associated privileges.  
Enforcement occurs as follows:

\noindgras{Receiving:~} If the receiving entity's labels do not agree with those of an attribute value, the attribute value (and any sub-attributes) are removed from (made {\small \textsf {null}} in) the message.
This is enforced on message receipt, before it is delivered to the entity.

\noindgras{Sending:~} An entity cannot send values for attributes where its labels do not agree with those of the attribute.
This is enforced when an entity attempts to send a message, ensuring values for any attributes violating this policy are removed, before message propagation.

Enforcement is automatic, meaning that applications using the messaging system can be subject to IFC enforcement completely transparently (\ie without their direct involvement); though again, there is the interface for the application to
actively manage IFC where required.
In addition, the general reconfiguration capabilities of the middleware enable connections between components to be defined and managed at runtime, providing another mechanism for controlling communication~\cite{debs14SBUStutorial}.

\subsection{Integrating With Persistent Storage}
\label{sec:os:service}

One technique to provide IFC with persistent data stores, 
is to store the tags alongside the data,
and a trusted software component ensures that when information is read from the store, the corresponding labels are applied. 
In Flume \cite{Krohn:2007:IFC:1294261.1294293}, a trusted process provides the interface between untrusted applications and persistent storage.
More recent work has seen the emergence of databases that natively understand IFC concepts and can enforce IFC policies \cite{Schultz:2013:IDI:2465351.2465357}. 

We 
see much promise in having the middleware mediate between persistence systems and the kernel, to ensure consistent IFC application.

\subsection{Evaluation}

\begin{figure}[t]
\centering
\begin{tikzpicture}
\begin{axis}[
    xbar stacked,
    legend style={
    legend columns=4,
        at={(xticklabel cs:0.5)},
        anchor=north,
        draw=none
    },
    ytick=data,
    axis y line*=none,
    axis x line*=bottom,
    tick label style={font=\footnotesize},
    legend style={font=\footnotesize},
    label style={font=\footnotesize},
    xtick={0,100,200,300,400,500,600,700},
    width=\columnwidth,
    bar width=3mm,
    xlabel={Time in ms.},
    yticklabels={},
    xmin=0,
    xmax=700,
    area legend,
    y=7mm,
    enlarge y limits={abs=0.625},
]
\addplot[ifc,fill=ifc] coordinates
{(71,0) };
\addplot[native,fill=native] coordinates
{(609,0) };

\legend{
IFC-overhead }\end{axis}  
\end{tikzpicture}
  \caption{IFC overhead of CamFlow-MW for a workload transmission of 5000 messages (x-axis in ms)     }
    \label{image:mw:eval}
\end{figure}
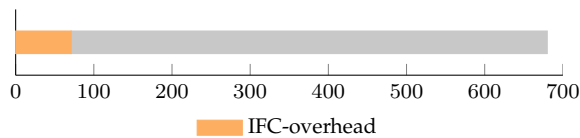

As shown in Fig. \ref{image:mw:eval},
the results indicate that IFC enforcement introduces an overhead of  $\sim$13\% in performance time compared to the standard, non IFC-enabled middleware (see \cite{singh2014ic2e} for details).
Note that these results were measured in the context of a particular workload, deliberately designed to highlight the impact of IFC enforcement. It follows that the overheads associated with real-world usage are most likely less onerous.

\section{Audit: Data-Centric Logs}
\label{sec:audit}

IFC, in addition to providing strong assurances that policy is being enforced, can also provide a data-centric log \cite{Ganjali:2012:ACM:2382536.2382549} detailing the information flows within and between system components. 
In addition to enforcing IFC via our LSM module we log the data flows of labelled processes, policy decisions, privileges and IFC security context manipulations.
Equivalent inter-machine operations via the middleware are also recorded.

Cloud logging systems are generally based on legacy logging systems (OS, web-server, database etc.) that either fail to capture the needed information, or are extremely complicated to interpret in a useful manner \cite{ko2011system}. More importantly, such logs tend to be relevant only to the particular service or component, which makes it difficult, if not impossible, to audit across a range of applications, clouds, etc.

IFC logs, as provided by our platform, allow us to capture information on application-level data flows, both attempted and permitted, allowing the correct expression and implementation of data flow policy to be checked. This provides transparency allows for meaningful audit, in terms of investigating the circumstances in which data leakage occurs, and provides evidence of compliance, \eg with legal obligations~\cite{singh2015:claw-magazine}.

\subsection{Analysing Paths to Disclosure}

\begin{figure}[t]
\centering
  \includegraphics[width=\columnwidth]{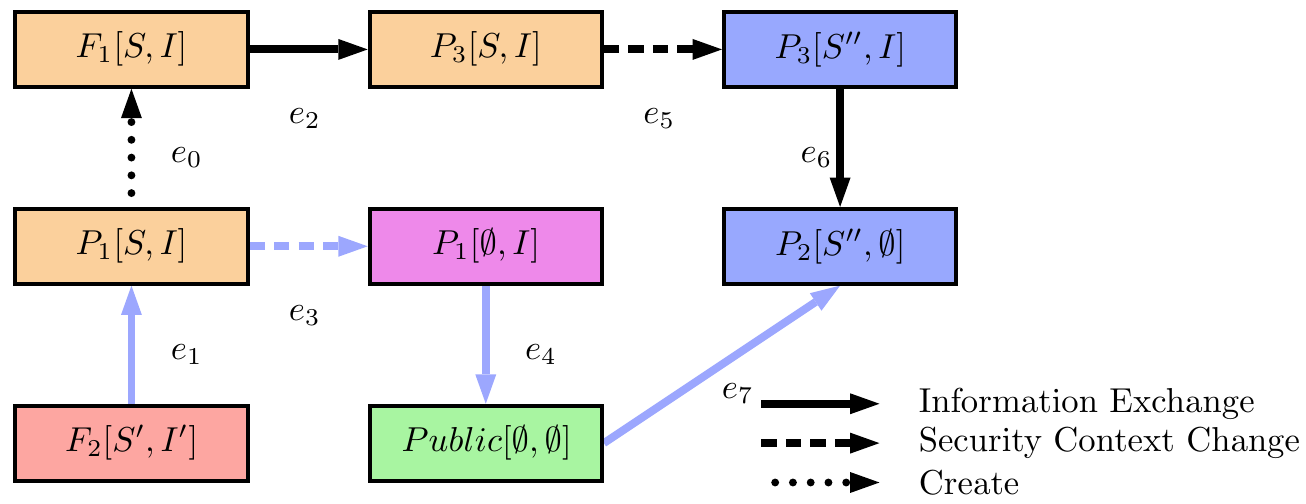}
  \caption{Simplified audit graph from IFC OS execution (we omit metadata for readability). The path to disclosure is shown in blue\slash pale.} 
  \label{image:audit:graph}
\end{figure}

To assist in interpreting log information, we build a directed graph corresponding to the allowed flows during the execution of our system, as shown in Fig. \ref{image:audit:graph}.
The flows defined in our IFC model (see \S\ref{sec:model}), namely \textbf{data flow}, \textbf{creation flow}, \textbf{security context} change and \textbf{privilege delegation}, correspond to the edges of the directed graph.
Entities (such as processes, files, messages \etc) are represented as nodes in the graph.

In addition to information necessary to build the graph (as shown in the figure), additional metadata is collected for forensic purposes, which is context/entity/event-dependent.
These audit entries, provided by the LSM and middleware, can be exploited by a dedicated service implemented in user space, connected to the kernel collection mechanism via \textsf{relayfs} \cite{zanussi2003relayfs}. 
This service feeds, for example, a graph visualisation tool such as Cytoscape \cite{smoot2011cytoscape} or a graph database such as Neo4J.\footnote{\url{http://neo4j.com/}}

Such a directed graph helps one identify data  leaks. For example, a tenant might discover that some sensitive medical data leaked into a data store where only anonymised research data were supposed to be stored. 
IFC is enforced in line with the policy encapsulated in labels; thus data may leak if such policy is improperly expressed and\slash or declassification\slash endorsement processes are not correctly implemented (\eg if the anonymisation process in Fig. \ref{image:declass} allows re-identification).

Suppose that an information leak is suspected between different security contexts $L_1[S, I]$ and $L_2[S', I']$.  Determining whether such a leak can occur is equivalent to discovering whether there is a path in the graph between the two contexts.
If the leak occurred, there must be a path between some entity $E_i$ such that $S(E_i)=S \wedge I(E_i)=I$ and another entity $F_i$ such that $S(F_i)=S' \wedge I(F_i)=I'$. 

The existence of such a path demonstrates that a leak is possible. To investigate whether a leak occurred, it is essential to consider the event ID associated with the edges comprising the path.
We denote by $e_i$, the last incoming edge to the entity under investigation with labels $[S', I']$; only edges such that $e<e_i$ should be considered.
When applied to all nodes along a path, this rule ensures strictly monotonically increasing timestamps from the first node to the last. 
Fig. \ref{image:audit:graph} shows in blue\slash pale a possible data disclosure path, from file $F_2$, from a very simple audit graph. We know from the event IDs $e_0$ and $e_1$ that the data disclosure did not occur through file $F_1$ and process $P_3$, but through $P_1$'s declassification.

\subsection{Demonstrating Compliance}

Compliance with certain requirements  can be demonstrated through queries over the graph.
We assume the audit data is stored in a graph database that we can query.
For example, the following plain English policy:\\
\emph{``European personal data sent to the US must be anonymised''}
\cite{pasquier2014.claw}, is equivalent to writing a query that verifies that there is no path between EU- and US-labelled data without an anonymisation process. \\
The policy \emph{``Medical data stored in database X must have received proper consent and be anonymised''} \cite{singh2015:claw-magazine} 
can be expressed as a query verifying that there is no path between data labelled as $\textsf{\small medical}$ and the database, without consent-checking and anonymiser processes. In addition, an investigator may want to know which anonymisation algorithm has been run, which data has been used to generate the anonymised records \etc 
Our audit graph assists in answering such questions.

Note that IFC only applies guarantees with respect to flows. Demonstrating the overall effectiveness of the management regime, \eg the quality and suitability of the anonymisation algorithm, is out-of-scope for a flow-based enforcement mechanism.

\subsection{Audit as `Big Data'}

We are potentially generating a vast amount of data in our IFC logs. 
However, unlike standard system logs that are complex to analyse, our logs generate graphs that are ideal for analysis by ``big data'' tools that have been developed for this purpose \cite{angles2008survey}.

Since the amount of data is potentially huge, the amount of data being logged can be fine-tuned to meet the requirements of the platform/tenant; e.g. 
by reducing the amount of metadata being stored, by logging only security context changing operations, by logging only information corresponding to some target security context, keeping operations on unlabelled entities outside of the log \etc
The decision on what needs to be logged then becomes a tradeoff between utility and the volume (cost) of log generated, which can be decided in order to correspond to legal or contractual requirements (for example, a regulated sector may need to have a fine-grained log to satisfy data forensic requirements). Indeed, as such an approach is new to the cloud, such considerations will be refined by experience, with best practices developing over time.

\subsection{Audit Access}

Logs can contain sensitive information and access to them should be controlled. 
This represents an area of our ongoing work.
 Traditional access controls clearly play a role; however, secrecy tags could also be leveraged.
For example, an auditor, before being granted access to audit logs, could be forced to demonstrate ownership of the corresponding secrecy IFC tags (for example through cryptographic means as in \S\ref{sec:tags:represent}).
The auditor may be granted access to a log entry only if $S(\mathit{origin})\cup S(\mathit{destination}) \subseteq S(\mathit{auditor})$.

\section{Example: Support for Web Services}
\label{sec:example}
\begin{figure*}[t]
\centering
  \includegraphics[width=\textwidth-6cm]{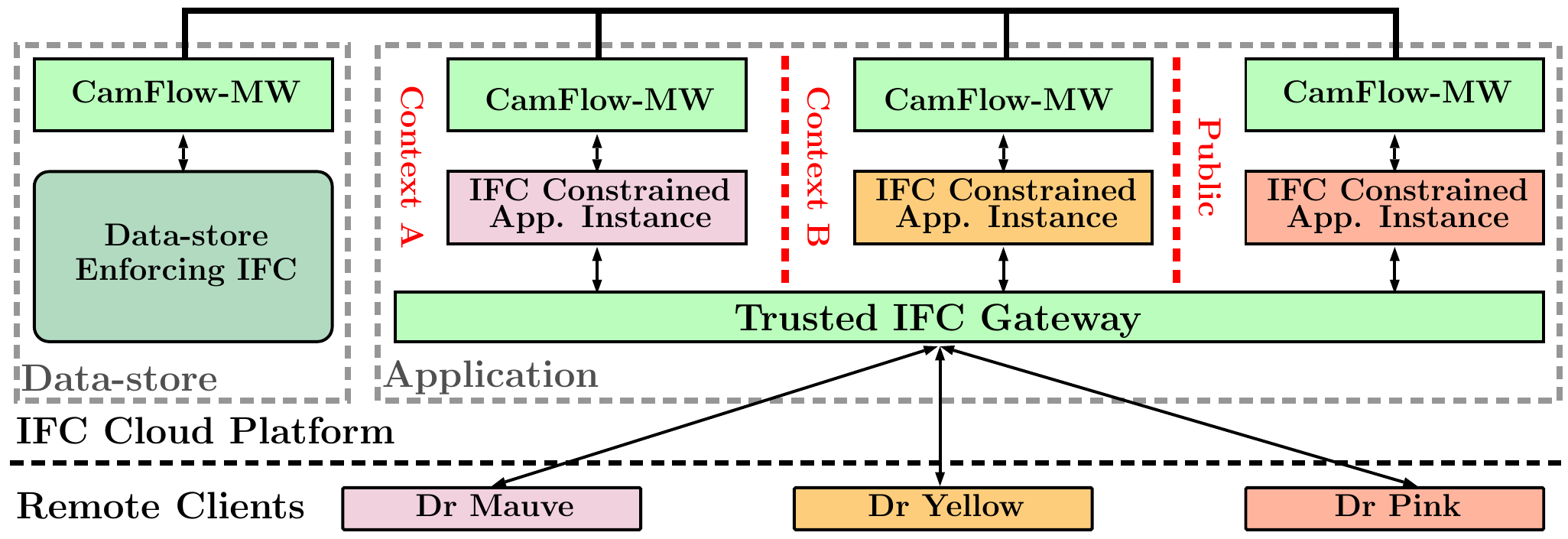}
  \caption{PaaS Architecture on top of IFC-OS}
  \label{image:untrusted:paas}
\end{figure*}

One of the most common uses of PaaS is to host web applications.
In this section we present the implementation of such a solution built on the infrastructure described in \S\ref{sec:os}, in order to evaluate and demonstrate the feasibility of our proposed approach. 
This is illustrated in  \fig \ref{image:untrusted:paas}.
We run standard and unmodified Ruby web applications.

Interaction with end-users
is achieved through a ``gateway'' between the IFC and non-IFC worlds.
Similarly, interaction with cloud services (such as data stores) is also achieved through our messaging middleware as discussed in \S\ref{sec:mw}. 
The requirement for this gateway can be removed if a trustworthy IFC implementation can be provided at the client side, consistent with the cloud implementation with respect to tag naming, enforcement, etc. 
Tag naming in general, system-wide, is an issue beyond the scope of this paper, see further \S\ref{sec:conclusion}. In our proof of concept implementation the gateway is a simple Apache server running a custom-built module.

The role of the gateway is to authenticate the end-user when a session is created, and to associate this session with an application instance running within the security context corresponding to the user. Recall that a security context comprises the $S$ and $I$ labels. Any further requests to the gateway in that session are routed to the corresponding application instance. Once an instance no longer has an associated session it can be recycled using self-checkpointing, as described in \S\ref{sec:os:checkpoint}.

Several application types are running over our cloud-based, web services platform. For example, in a medical context these might be medical record editing, pharmacy ordering, social services etc.
A single, shared, identity service for the end-user is part of the cloud provider offering (in our proof of concept implementation we used OAuth\cite{hardt2012oauth}).

The GP authenticates, is authorised as treating doctor for Alice and selects the `identity' that corresponds to Alice. A new session is created server-side by the gateway, 
with the requested application instance running in the corresponding security context, 
with $S=[\textsf{\small medical, alice}], I=[\emptyset]$.
When the GP wants to access applications on behalf of a new patient, he needs to close Alice's session, authorise as treating doctor for Bob and open a new session for Bob.

The control described above is not achieved by the application, but by the platform itself and can be controlled by the end-user, subject to access control. 
That is, a medical application used on Alice's behalf runs in a security context in which data cannot flow to that of another patient. 
Furthermore, applications running on behalf of a given user can share the data of that user without the risk of seeing a buggy application leaking data between end-users.

As described in \S\ref{sec:os}, we assume the middleware and the OS enforcement are provided as a service by the underlying platform. 
A tenant wanting to use the third-party, web-service offering, once his trust in the underlying platform is established, needs only to audit the gateway; again, the underlying infrastructure provider could either provide such a gateway or audit it.
The rest of the software stack of the third-party, web-service provider is bound by the IFC enforcement mechanism and therefore need not be trusted.

\section{Conclusion \& Future Work}
\label{sec:conclusion}

IFC  allows data flows to be controlled continuously throughout a system, by providing an information-centric MAC scheme that continuously ensures non-interference between security contexts.
This paper presented the CamFlow platform, that demonstrates the potential of cloud-deployed IFC as supporting: 
(1) protection of applications from each other; 
(2) flexible data sharing across isolation boundaries;
(3) prevention of data leakage due to bugs/misconfigurations;
(4) extension of access control beyond application boundaries;
(5) data flow transparency.

Specifically, we detailed a new kernel implementation of IFC as an LSM, demonstrating low overhead even for worst-case scenarios, where processes continuously make read/write system calls.  We also described the integration of a messaging middleware to enforcing IFC across machines.
This combination makes it possible to provide whole-system IFC to PaaS cloud services, and therefore also SaaS.
Our approach and implementation were designed so that applications can run unchanged over IFC, thus making cloud adoption feasible.

We also indicated how the data-centric logs based on IFC enforcement could provide the means to audit an IFC-enabled system, whereby a log can be processed as a directed graph to investigate leaks and attacks and show compliance with data management requirements. Though this represents our initial work in the area, there appears much promise. 

In light of the above, we believe that IFC has great potential as a security mechanism for the cloud whereby trust in a few major cloud providers, deploying IFC, can be built on to provide a demonstrably trustworthy computing environment.

CamFlow was developed with cloud deployment in mind. 
Our future work will investigate the challenges of a broader distributed context.

It is already feasible to extend CamFlow to support mobile environments. Android 
supports the full SELinux enforcement,\footnote{\url{https://
source.android.com/devices/tech/security/selinux}} and  an 
Android-LSM integration has been demonstrated~\cite{smalley2013security}.
But when dealing with multiple cloud services, particularly as they become part of a wider distributed architecture such as in the IoT, a trustworthy, system-wide deployment of IFC can no longer be assumed.
Much work remains on establishing trust in (and the trustworthiness of) the IFC enforcement mechanism within end-users' devices.
Outside a cloud context, all parties' trust in a common third party's enforcement of IFC constraints, cannot be assumed (unlike the cloud provider for cloud services).
We intend to explore leveraging hardware roots of trust and remote attestation to ensure the integrity and trustworthiness of IFC enforcement mechanisms.

Another area of investigation concerns the representation of tags across administrative domains.
In a cloud context, a federated approach can be envisaged where a common understanding of tags could be negotiated across multiple domains.
For example, in \cite{singh2015:claw-magazine} we discussed initial thoughts on managing data according to specific obligation regimes.
However, as the number of administrative domains increases, both a global tag naming scheme and mechanisms for ad-hoc negotiation become necessary.

Related is the sensitivity of the tags themselves.
Knowledge of the meaning of a tag can indicate that the associated entity contains or deals with certain information.
If a relationship can be established between an entity and the information owner, this may disclose private information about the information owner.
This may lead to work on a need-to-know negotiation mechanism to establish a secure channel between hosts, especially in a wide-scale distributed system.
A promising mechanism is \emph{private set intersection} to determine tags' subset relationship (\S\ref{safemsg}).

Another challenge concerns extending audit to distributed architectures, both in terms of resource management  and regulating access to log data.

\ifCLASSOPTIONcompsoc
    \section*{Acknowledgments}
\else
    \section*{Acknowledgment}
\fi
\noindent This work was supported by UK Engineering and Physical Sciences Research Council grant EP/K011510 CloudSafetyNet: End-to-End Application Security in the Cloud. We acknowledge the support of Microsoft 
through the Microsoft Cloud Computing Research Centre. 
 
\bibliographystyle{IEEEtran}
\bibliography{./biblio}
\end{document}